# Innovative remotely-controlled bending device for thin silicon and germanium crystals

**D. De Salvador**[a], **S. Carturan**[a], **A. Mazzolari**[b,1] **E. Bagli**[b], **L. Bandiera**[b], **C. Durighello**[a], **G. Germogli**[b], **V. Guidi**[b], **P. Klag**[c], **W. Lauth**[c], **G. Maggioni**[a], **M. Romagnoni**[b], **A. Sytov**[b]

[a] *INFN INFN Laboratori Nazionali di Legnaro, Viale dell'Universita 2, 35020 Legnaro, Italy*
*Dipartimento di Fisica, Universita di Padova, Via Marzolo 8, 35131 Padova, Italy*

[b] *INFN Sezione di Ferrara, Dipartimento di Fisica e Scienze della Terra, Universita di Ferrara Via Saragat 1, 44100 Ferrara, Italy*

[c] *IInstitut f• ur Kernphysik der Universit• at Mainz, D-55099 Mainz, Germany*
   E-mail: mazzolari@fe.infn.it

ABSTRACT: Steering of negatively charged particle beams below 1 GeV has demonstrated to be possible with thin bent silicon and germanium crystals. A newly designed mechanical holder was used for bending crystals, since it allows a remotely-controlled adjustment of crystal bending and compensation of unwanted torsion. Bent crystals were installed and tested at the MAMI Mainz MIcrotron to achieve steering of 0.855-GeV electrons at different bending radii. We report the description and characterization of the innovative bending device developed at INFN Laboratori Nazionali di Legnaro (LNL).

---

## Contents

## 1. Introduction

Experimental studies of manipulation with bent crystals via channeling [1] and Volume-Reflection [2] effects received a strong boost in recent years due to technological development which led to the successful demonstration of the LHC 6.5 TeV proton beam deflection [3].

The steering of electrons and negatively-charged particles in general is more challenging due to the shorter dechanneling length compared to their positive counterpart. Indeed, being attracted toward positive nuclei, channeled negatively-charged particles oscillate around crystal planes and not between them as it is for positive ones. For this reason, they are more affected by incoherent scattering with nuclei that causes a faster dechanneling of the particle trajectory. This fact required the development of thinner crystals [4] to overcome the problem of a shorter dechanneling length and a deeper understanding of scattering processes involved, thus leading to important results in beam steering [5,6,7], e.m. radiation generation in bent crystals [8,9] and the observation of quasi-channeling oscillation [10,11].

In order to exploit channeling phenomena of negative particles high quality bent crystals has to be produced down to few tens of micrometers.The most used materials of are Si or Ge that have been developed with high lattice qualities for micro-electronics applications. The mechanical bending of such brittle materials [12] is a non-trivial task especially if thin crystal for negative particle application has to be developed. The reason is that in usual bending systems holder the primary curvature is given during the mechanical clamping procedure causing local stress concentrations that may break the sample during mounting [13].

In the present paper we present an innovative bending scheme to be applied to thin (down to 15 micron or possibly less) silicon or germanium slabs. Samples are clamped when they are flat before the bending in order to minimize local stress. Then the samples are gradually and homogeneously bent without stress concentration, by actuating a single translational degree of freedom. A further degree of freedom to tune the curvature quality is also present and the possibility of manage chemical thinning procedure after clamping and before bendig are demonstrated.

Examples of deflected channeling distributions of 855 MeV electrons available at the MAMI B line obtained with a 15 μm Si crystal bent through the the piezo-bending device are finally showed and experimental details can be found in [14].

## 2. Bending device

### 2.1 General features

The basic principle is that of the bow: the crystal slab sides are put under tension by squeezing the sides one toward the other by a translation movement. To avoid stress at the sides, the slab is clamped on two freely rotating surfaces. During bending the maximum stress is obtained in the middle of slab while it goes to zero close to clamping. In this way low stress concentration occurs at clamping point and the specimen can be used for the application up to curvatures close to the breaking limit.

In this paper we describe the design, the realization and the use of a bending device prototype specifically developed for channeling application, in particular the system has been used to deflect 0.855 GeV electrons by means of planar channeling into (111) Si bent planes at the MAMI microtron of Mainz with an experimental setup based on the one presented in Ref. [15]. For this application the quality of the bending is mandatory since of the very limited acceptance angle over which the channeling phenomenon is efficient. Spurious bending (torsion) due to mechanical misalignment are compensated by an additional degree of freedom that takes care of the parallelism between the rotating clamping planes [16]. The torsion correction reduces spurious stress contributing to the realization of high curvatures. The squeezing and torsion degrees of freedoms are remotely controlled by piezo motors and the system can operate in high vacuum needed for the electron beam transport. The system allows to adjust the curvature and the torsion at very high precision without breaking the vacuum with strong advantage for the data taking speed.

Moreover the bending scheme can be used to obtain very thin bent specimens. Even if the clamping on flat surfaces before the bending is a strong advantage, the specimen can be hardly managed and often break during the clamping procedure if they are below certain thickness (10-15 micron for silicon, 20-30 for the more brittle germanium). In order to overcome this limit too, the clamping is performed with sample over the critical thickness and the specimen are further thinned by chemical procedures before being transferred to the bending device. In summary the procedure is performed in the following steps: i) clamping of the specimen to the rotation axis; ii) Eventual further thinning by chemical procedures; iii) transfer to the bender; iv) bending of the specimen. With this method down to 15 μm Si and Ge are bent down to 3 mm curvature radius.

### 2.2 Design description

The design of the bending system is reported in Fig.1. The bender is made by different mechanical parts (A to G) including two piezo motors (D: Picomotor actuator 8301-UHV, F: Newport AGLS25). Specimen (H) is clamped to two plugs I1 and I2 that can be transferred into the bender with a translation along axes indicated with black dashed lines. During transfer (Fig 2.a) the plugs are mechanically fixed to a handling holder that avoids the accidental rotation of the plugs and allows to handle also very thin and fragile samples. After complete transfer the handling holder is removed and the plugs are inserted in two couples of holes where they can freely rotate. In order to obtain the sample bending, part A is translated toward part B in order to reduce the distance between the plugs. During translation plugs rotates to minimize the stress of the sample providing a homogenous bow-wise curvature of the specimen. The translation is finely controlled by the linear translation stage F with a step size of about 0.5 μm.

In principle this single translational degree of freedom may be used in order to tune the desired curvature. Practically the obtained curvature could be not satisfactory since of possible imperfection in the parallelism of the two clamped portions of the sample. In this case additional unwanted deformations occurs that may change the channeling alignment over the beam size thus reducing the efficiency. Lack of parallelism may be induced by mechanical imperfection in the clamping procedure (that is actually performed by double adhesive kapton tape), by mechanical imperfection of the holder or by spurious rotations of the translation stage during movement. An additional remotely controlled degree of freedom is therefore included into the device. It allows to rotate one of the plugs with respect to the other correcting parallelism. The plug holder B can rotate with respect to the fixed part E around a cylindrical plug C with a rotation axis parallel to the y direction. The cylindrical plug C is only partially visible in Fig.1 since both B and E parts have cylindrical hollows that act as a guide for the rotation. The rotation regulation occurs by a push-pull system: the linear actuator D pushes the plug holder B while, on the opposite side (not visible), a spring pulls and keeps stability. This trick allows to remotely control the parallelism of I1 and I2 for rotation around the y-axis, by the remote control of the linear piezo motor D. Given the fine step of about 30 nm and the lever distance of the push pull system (20 mm), the system can in principle regulate the parallelism with a resolution lower than 1 µrad.

Such fine tuning is useful for regulating the torsion of the specimen that is a typical problem for channeling application [17]. If the parallelism is not finely tuned, the specimen undergoes a screw like deformation detrimental for the beam alignment with channeling planes. As a consequence, only a fraction of the beam would result to be channeled and deflected by the device.

The push pull system regulate the orientation of I2 with respect to I1 by means of a rotation around the y axis. In order to obtain a full parallelism between the supporting plugs a rotation around z-axis should be applied to one of them. To this aim, mechanical part A is machined with a symmetric regress that allows to elastically rotate the support of I1 by means of a push pull action. Push pull is actuated by two non-motorized screws and is calibrated before the sample transfer by using a parallel x-ray beam and a high precision goniometer to check the parallelism.

The holder is thought to operate with the vertical direction along the x-axis as shown in Fig.2b. In this case the I plugs are avoided to fall by a screw with spherical head centered in the rotation axis (G in Fig.1). The friction momentum due to the I plugs weight is strongly minimized since they have to virtually rotate around a point. During the bending the specimen stress slightly push the plugs against their hole walls causing friction. This effect is minimized by using rectified stainless-steel plugs and brass hole holder A and B.

The suitable hole plugs coupling and the whole mechanical accuracy is guaranteed by high precision wire electrical discharge machines used for fabrication.

## 2.3 Sample preparation and bending

Both Si and Ge specimen were produced and bent through the piezo-bender. Silicon samples with tens of microns thickness are produced as shown in ref. [4].

Germanium thin plates were produced starting from a 4 inch diameter wafer purchased from Umicore (Olen, Belgium), <211> orientation, EPD lower than 1000 /cm$^2$ and thickness ranging from 30 to 70 µm as claimed by the factory. A dedicated procedure to cut the wafer by chemical etching has been followed in order to avoid breakage by firstly depositing by RF sputtering a pattern of protecting Cr/Au pads with size 15×13 mm$^2$ on both sides of the wafer, through a mechanical mask. Then, the wafer was immersed in an etching bath composed of nitric acid (HNO$_3$, 65%) and hydrofluoric acid (HF, 50%) 3:1 volume ratio for few minutes, as predicted by the etch rate of around 20 µm/min [18].

The Cr/Au coating was then stripped from the surface of the so obtained rectangular plates by using a gold etchant bath based on a solution of potassium iodide/iodine (Gold etchant, Sigma Aldrich) and concentrated hydrochloric acid (HCl, 37%) to remove Chromium.

Further wet chemical processing has been applied to thin the plate down to around 15 µm in thickness by using a low rate etching bath, based on hydrofluoric acid, hydrogen peroxide ($H_2O_2$, 27%) and bidistilled water 3:3:94 volume ratio [19]. The etch rate and the homogeneity in erosion of this solution have been characterized by means of HR-XRD, exploiting the Beer-Lambert law [20], and the final rate of 0.78 ± 0.07 µm/min has been derived. After a first thinning step in this bath, the Germanium plate-like crystal has been anchored using two-side Kapton adhesive tape to the holder plugs and underwent a further thinning step on board. Then, a final thickness characterization has been performed and the collected XRD map is reported in Figure 3.

Crystals are clamped to the plugs, transferred to the holder and bent. In Fig.2a a bent silicon crystal with a 15 micron thickness is shown. The sample of the photo has a visible primary curvature along the main surface, which is parallel to the (211) planes, with a radius of 2.8 mm. A secondary bending of 10 mm of the (111) inner planes due to the quasi-mosaic effect [14,21] is exploited for channeling.

**2.4   Experimental test**

Before measurements, torsion correction is performed: in Fig.4 a set of data demonstrating the procedure are shown. The channeling angle is obtained by aligning the Si bent crystal with respect to a small electron beam using the MAMI B line set-up [15]. In the reference frame of Fig.1, the beam reaches the sample along z-axis and its footprint in the vertical x-y plane is 0.1 mm x 0.2 mm root means square. Channeling angle of (111) planes of the crystal, parallel to y-z plane, is determined by rotating the whole holder around y-axis. The alignment is determined with the beam on the middle of the sample and after translation of the holder of +1 or -1 mm along the y axis (lateral position in Fig.4). Measurements are then repeated by varying the stroke of the motorized screw D. At initial position (stroke 0 µm) the crystal channeling alignment change with lateral position. This is due to the fact that I1 and I2 plugs are not parallel causing a torsion of the sample. Torsion can be quantified by the slope of the linear fit of the data and turn out to be 900 µrad/mm before acting the screw. By moving the D screw, the data becomes flatter and flatter reaching a final torsion value of 35 µrad/mm after a stroke of 145 µm. This value is satisfactory since it means that a misalignment of 7 µrad is now present along the beam size of 0.2 mm. This is completely negligible considering that the critical angle of Si (111) is about 200 µrad for 855 MeV electrons of MAMI. It is worth to note that the D screw allows for much finer steps (about 30 nm) and in principle much better correction could be reached in case of more critical situations as it occurs at higher beam energies with lower critical angle.

An experiment has been carried out at the MAinzer MIkrotron B line with 855 MeV electrons to measure the steering capability of 15 µm long silicon and germanium crystals, bent along (111) planes. The usage of the piezo-actuated mechanical holder allowed to remotely change the crystal curvature, it was possible to study the steering capability of planar channeling vs. the curvature radius. For silicon, the channeling efficiency exceeds 35 %, a record for negatively charged particles. This was possible due to the realization of a crystal with a thickness of the order of the dechanneling length. On the other hand, for germanium the efficiency is slightly below 10 % due to the stronger contribution of multiple scattering for a higher-Z material. Nevertheless this is the first evidence of negative beam steering by planar channeling in a Ge crystal. Details about experimental procedure and results on deflection efficiency and de-channeling length vs. the curvature radius for both Si and Ge crystals are published in [14]. Fig. 5 shows the experimental distributions of the deflected 0.855 GeV electron beam as measured with a LYSO screen at MAMI for 5 different curvature radii.

## 3. Conclusions

In this paper an innovative remotely controlled crystal bender system is presented. The bending device demonstrates many advantages: The clamping of the sample is performed on flat surfaces before impressing the curvature, this reduces stress concentration and allows to perform clamping with a reasonable success rate of Si samples down to 15 micron thickness and Ge sample down to about 25 micron. The clamped sample can be further thin down on-board before bending. This procedure was successfully applied to Ge slab in order to reduce the thickness down to about 15 micron, and could in principle be applied to get even thinner Si sample. After transfer the holder impress a bow-like primary curvature to the sample up to impressive value of 3mm, by means of one single remotely controlled motor. Curvature may be changed without vacuum breaking. The curvature quality is regulated by a further piezo-motor that can very finely regulate the torsion. Data demonstrating this function are illustrated. Finally, channeling deflection spectra are shown as an example, to complement the data already reported and analyzed in ref. [14].

The present device is also an innovative proof of concept that could be scaled and optimized for different sample thickness to satisfy channeling requirements at different beam energies. Moreover, similar device may be also developed for x-ray beam manipulation applications.

## 4. Acknowledgments

We acknowledge partial support of INFN CSN5 AXIAL experiment and of the European commission under the H2020 PEARL project (GA n.690991) and FP7 CRYSBEAM project (GA n.615089). We acknowledge M. Rampazzo, A. Minarello e A. Pitacco for technical work on the bending system project and realization.

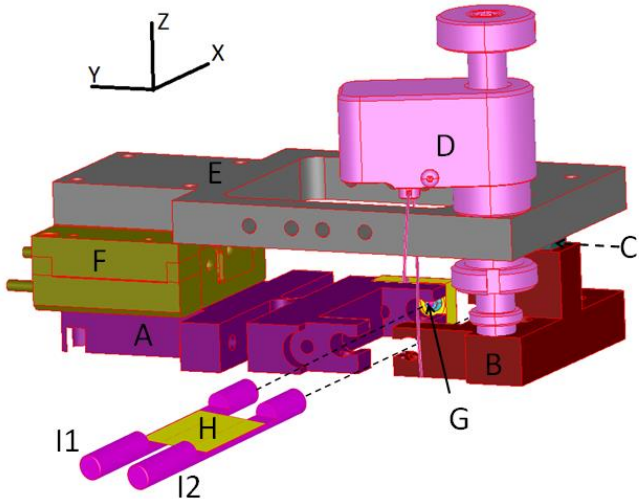

Fig.1 Ensemble drawing of the bending system

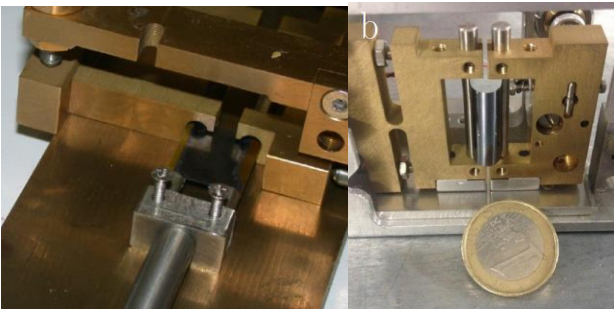

Fig.2 a) Sample transfer operation. b) Bent Si sample

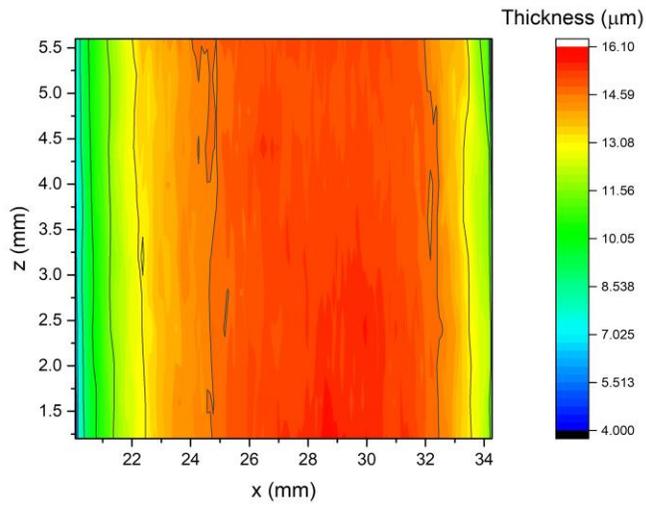

Fig.3 XRD derived thickness bidimensional map of the Ge plate-like crystal.

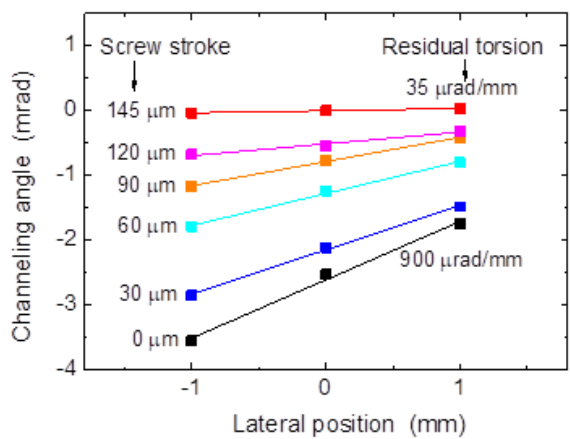

Fig.4 Channeling angle of (111) plane by changing the lateral position and the D screw stroke. The screw movement allows to effectively correct the torsion.

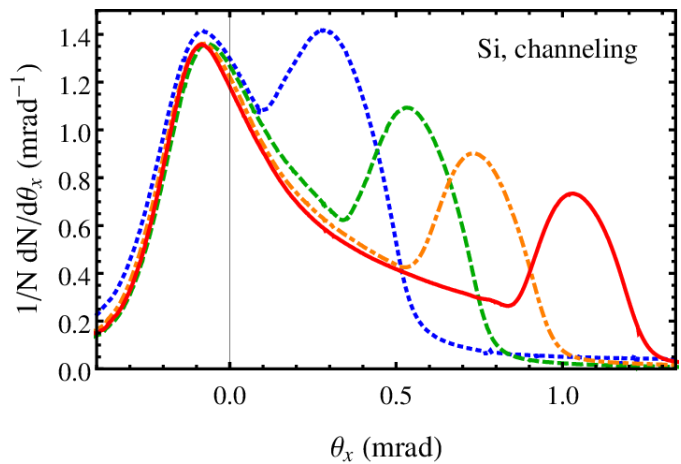

Fig.5 Deflected 0.855 GeV electron beam distribution by channeling in the 15 μm long Si crystal bent through the piezo-holder. Four different quasi-mosaic curvature radii, $R$, were used: 47.5 *mm* (blue curve), 27.2 *mm* (green curve), 20 *mm* (orange curve) and 13.9 *mm* (red curve), respectively.